\documentclass{article}
\usepackage{caption}
\usepackage[utf8]{inputenc}
\usepackage{amsmath,amsthm,bm,mathrsfs,mathtools, hyperref}
\usepackage{mdframed}
\usepackage{authblk}
\hypersetup{
    colorlinks,
    citecolor=black,
    filecolor=black,
    linkcolor=black,
    urlcolor=black
}
\usepackage[a4paper, total={7in, 10in}]{geometry}
\usepackage{graphicx, float}
\usepackage[version=4]{mhchem}
\usepackage[sorting=none]{biblatex}
\title{On the Origins of Life’s Homochirality: Inducing
Enantiomeric Excess with Spin-Polarized Electrons}

\author[1]{S. Furkan Ozturk}
\author[2]{Dimitar D. Sasselov}

\affil[1]{Department of Physics, Harvard University, Cambridge,
Massachusetts 02138, United States}
\affil[2]{Harvard-Smithsonian Center for Astrophysics, Cambridge, MA 02138, United States}

\date{\today}

\begin{document}
\maketitle

\begin{abstract}
    Life as we know it is homochiral, but the origins of biological homochirality on early Earth remain elusive. Shallow closed-basin lakes are a plausible prebiotic environment on early Earth, and most are expected to have significant sedimentary magnetite deposits. We hypothesize that UV (200-300nm) irradiation of magnetite deposits could generate hydrated spin-polarized electrons sufficient to induce chirally selective prebiotic chemistry. Such electrons are potent reducing agents that drive reduction reactions where the spin polarization direction can alter enantioselectively the reaction kinetics. Our estimate of this chiral bias is based on the strong effective spin-orbit coupling observed in the chiral-induced spin selectivity (CISS) effect, as applied to energy differences in reduction reactions for different isomers. In the original CISS experiments, spin selective electron transmission through a monolayer of dsDNA molecules is observed at room temperature\textemdash indicating a strong coupling between molecular chirality and electron spin. We propose that the chiral symmetry breaking due to the CISS effect, when applied to reduction chemistry, can induce enantioselective synthesis on the prebiotic Earth and thus facilitate the homochiral assembly of life's building blocks.
\end{abstract}

\begin{mdframed}
\section*{Significance Statement}
Essential biomolecules like amino acids and sugars are chiral\textemdash they exist in mirror symmetrical pairs named enantiomers. However, modern life selectively uses only one of the enantiomers. The origin of this chiral symmetry breaking remains elusive to date and is a major puzzle in the origin of life research. Here, we consider spin-polarized electrons as potential chiral symmetry breaking agents utilizing the robust coupling between electron spin and molecular chirality at room temperature as established by the CISS effect. We propose that chiral bias is induced and maintained with enantioselective reduction chemistry driven by such spin-polarized electrons that are ejected from magnetite deposits in shallow prebiotic lakes by solar UV irradiation.
\end{mdframed}

\section{Introduction}

Chirality (or handedness) is a geometric property, and a molecule that cannot be superimposed on its mirror image is said to be chiral \cite{kelvin1904baltimore}. Molecules with opposite handedness (called enantiomers) show identical chemical behavior, although biology is picky when it comes to chirality. Essential molecules for life: amino acids and sugars are selectively used in only one handedness. All biological systems use left-handed (L) amino acids and right-handed (D) sugars. Therefore, homochirality is considered to be a signature of life, and understanding the origins of homochirality is essential for understanding the origins of life. However, the origins of biological homochirality remain an open problem. Nonetheless, it is often acknowledged that its early emergence, e.g., during the prebiotic synthesis of the monomers, would be very advantageous in achieving the efficient polymerization of RNA, which is inhibited in a racemic mixture of nucleotides \cite{joyce1984chiral, blackmond2010origin}.

Reaching a homochiral state requires at least two things. First, a chiral symmetry breaking agent that can induce an enantiomeric excess (ee) (\%ee = 100 * (L-D)/(L+D)). Second, a prebiotically plausible mechanism that can amplify this imbalance \cite{blackmond2010origin}. Regarding the latter, a number of asymmetric amplification mechanisms have been proposed. The Soai group has reported that asymmetric autocatalysis can generate nearly perfect ee with high yields\textemdash although the conditions are not prebiotically plausible \cite{shibata1999practically}. Blackmond and co-workers \cite{hein2011route} demonstrated that RNA precursors can be enantioenriched by chiral aminoacids in the Powner-Sutherland ribonucleotide synthesis \cite{powner2009synthesis}. Their scheme showed amplification in the ee of pyrimidine nucleotide precursors such as glyceraldehyde and amino-oxazolines. The Blackmond group also showed that chiral pentose sugars can chirally enrich aminoacid precursors with substantial amplification \cite{legnani2021mechanistic}. Their analysis revealed a significant dynamic kinetic resolution that can take advantage of the selective reaction rates for enantiomers and amplify a small ee into near unity. In combination, these results indicate that sugars can trigger the enantioenrichment of aminoacids and vice versa, and inducing a small chiral excess can be sufficient to reach a chirally pure state. However, the search for a prebiotically plausible asymmetric amplification mechanism is still active \cite{blackmond2019autocatalytic}. In this work we are investigating a symmetry breaking agent that can trigger such an amplification mechanism.

\begin{figure}[h]
    \centering
    \includegraphics[width=\textwidth]{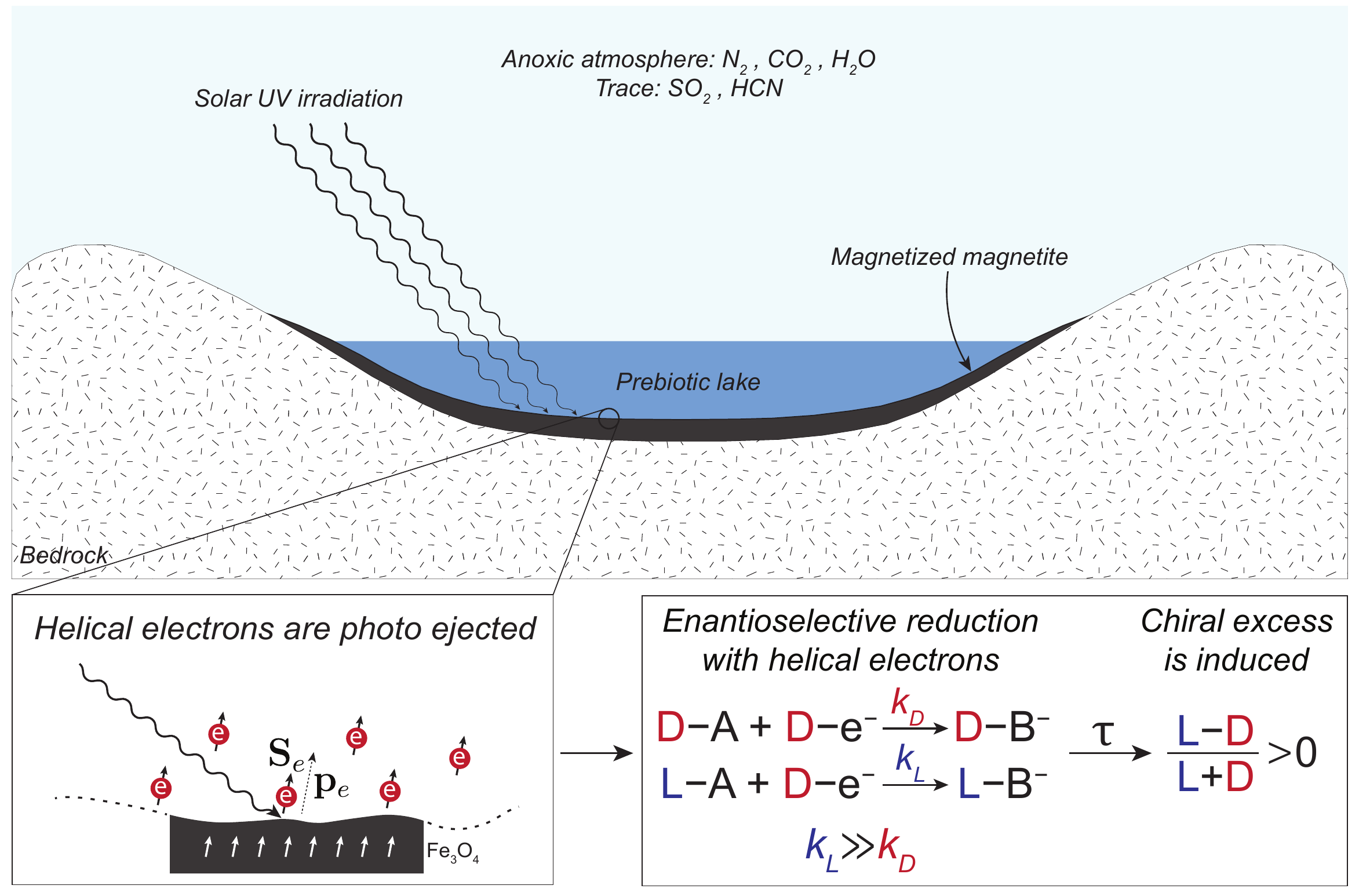}
    \caption{An evaporative lake with magnetite deposits contains the feedstock molecules for prebiotic chemistry. Irradiation of the uniformly magnetized magnetite (\ce{Fe3O4}) deposits with solar UV ($200$-$300$nm) light generates helical photoelectrons. Helicity of the electrons (\ce{D-e^-} in the figure) is determined by the magnetization direction. Helical electrons induce CISS-driven reduction chemistry (CDRC) near the magnetite surface due to a selectivity in the reaction rates, $k_L$ and $k_D$, for different isomers L and D. This selectivity in the reaction rates can induce an imbalance between two isomers. In the figure, chiral excess in the L isomer is induced.}
    \label{Fig.1}
\end{figure}

Many chiral symmetry breaking agents have been proposed\textemdash CPL, magnetic fields, cosmic rays, and weak nuclear currents to name a few \cite{balavoine1974preparation, noorduin2009complete, rikken2000enantioselective, globus2020chiral, szabo1999demonstration, mason1987universal}. These symmetry breaking agents induce ee either by selectively producing or destroying one isomer more than the other. Studies with CPL take advantage of the optical activity of chiral molecules and it has been shown that a nearly homochiral state is reached with the amplification of a slight chiral excess induced by CPL \cite{balavoine1974preparation, noorduin2009complete, shibata1998amplification}. However, the efficiency of CPL is low and the source and availability of CPL on the early Earth is unclear. Utilizing optical activity of chiral molecules, albeit using unpolarized light in an external magnetic field, Rikken and Raupach demonstrated that a chiral excess of about $100$ ppm can be generated \cite{rikken2000enantioselective}. Though this so-called magnetochiral excess is realized with unpolarized light, it relies on the presence of external fields as high as 10 Tesla. Polarized cosmic rays have been theorized as a universal enantioselective agent inducing a bias during the evolution of two biosystems with opposite chirality \cite{globus2020chiral}. Experimental support has come from enantioselective destruction due to dissociative electron attachment (DEA) with low-energy longitudinally polarized electrons by Dreiling and Gay \cite{dreiling2014chirally}, highlighting the viability of longitudinally polarized cosmic beta radiation to do that. Prior experiments by Rosenberg et al. had presented a different version of DEA and demonstrated chirally selective destruction of a racemic adsorbed layer of chiral molecules, yet with an alternative chiral symmetry breaking agent: spin-polarized low-energy (a few eV) secondary electrons \cite{rosenberg2008chiral}. The effect they observed is the enantioselective destruction of a chemical bond via high-energy scattering under ultra-high vacuum conditions and it appears to be due to a strong coupling between the spin-polarized electrons and the molecular chiral center \cite{rosenberg2008chiral, rosenberg2015chiral}. 

We suggest that the same strong coupling induces chiral selectivity, but with a different source and mechanism. We propose an enantioselective reduction chemistry, in solution, induced by spin-polarized photoelectrons ejected from a magnetized surface, as shown in Figure 1. The mechanism we suggest is yield-preserving as it drives enantioselective production, not destruction. It operates under conditions that are compatible with the conditions of a reduction chemistry at room temperature, in solution. The core of our mechanism is the spin-chirality interaction. But how does spin interact with molecular chirality and induce enantioselective chemistry?

\section{The CISS Effect - Chiral Molecules and the Electron Spin}

\begin{figure}[h]
    \centering
    \includegraphics[width=\textwidth]{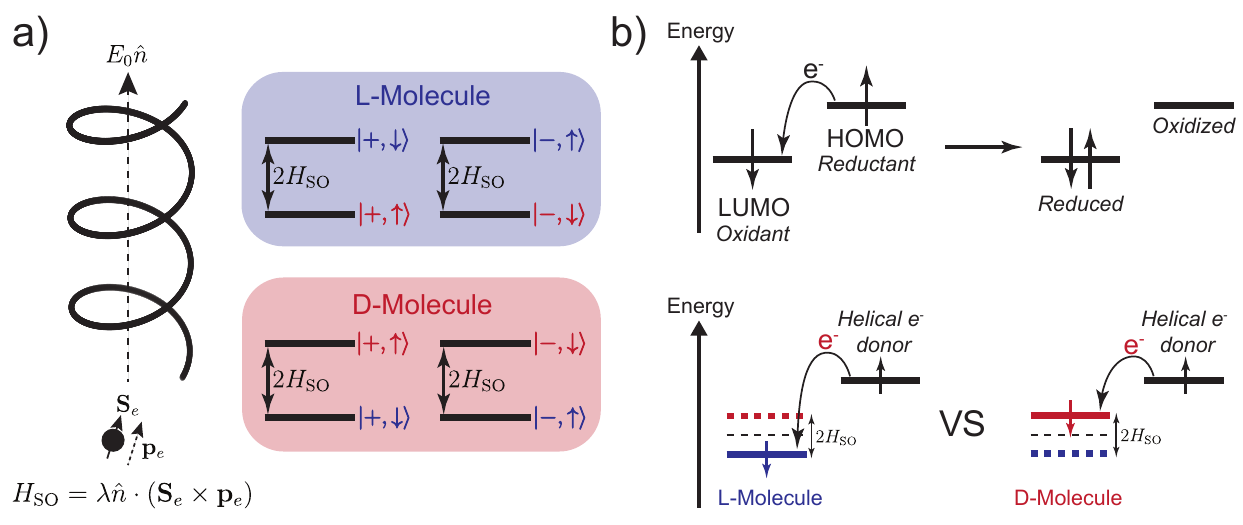}
    \caption{a) The CISS effect strongly couples molecular chirality and electron spin. Electrons interacting with a chiral molecule are spin filtered based on the relationship between their spin and momentum directions. L-molecules energetically prefer $|D\rangle$ electrons and vice versa and the energetic difference between two helicity states is given by the effective spin-orbit energy $2H_{\text{SO}}$ b) The driving force of a reduction reaction is the energy difference between the molecular orbitals of the reductant and oxidant. The CISS effect causes an energy difference of molecular orbitals for different isomers when a reduction reaction is driven by helical electrons. An isomer with LUMO at a lower energy (L in the figure) is reduced faster and this causes enantioselectivity due to the differing reaction rates.}
    \label{Fig.2}
\end{figure}

Since the observation of the chiral-induced spin selectivity (CISS) effect the strong coupling between electron spin and molecular chirality has taken front stage \cite{ray1999asymmetric}. The initial experiments by Naaman and co-workers showed that a self-assembled monolayer of dsDNA molecules spin filter a current of photoelectrons on a gold surface \cite{gohler2011spin}. Later studies have confirmed that the effect is induced by the molecular chirality and is robust at room temperature \cite{naaman2019chiral, gohler2011spin}. Although a full theoretical framework is still missing, CISS phenomena are qualitatively explained by the coupling of the electron's linear momentum ($\mathbf{p}_e$) and spin ($\mathbf{S}_e$) in the presence of a chiral electrostatic potential $\textbf{E}_\text{chiral}$ that changes sign when the handedness is reversed \cite{naaman2012chiral}. This spin-orbit effect is due to an effective magnetic field, $\textbf{B}=-\frac{1}{c^2}\textbf{v}\times\textbf{E}_\text{chiral}$, the electron experiences in its rest frame as it goes through a chiral molecule,  where $\textbf{v}$ is the electron velocity, and $c$ is the speed of light. The electron interacts with this effective magnetic field through its spin magnetic moment $\pmb{\mu}=-g \mu_B \textbf{S}_e/\hbar$, where g ($\approx2$) is the spin g-factor, $\mu_B$ is the Bohr magneton, and $\hbar$ is the Planck constant. This interaction leads to a splitting in the electron energy as represented by the spin-orbit Hamiltonian $H_{\text{SO}}=-\pmb{\mu}\cdot\mathbf{B}$.

\begin{equation}
    H_{\text{SO}} = \lambda\hat{n}\cdot\left(\mathbf{S}_e \times \mathbf{p}_e \right)
\end{equation}

where the coefficient $\lambda$ is defined as $\lambda \equiv \frac{E_0 g\mu_B}{\hbar m c^2}$ for $\textbf{E}_\text{chiral} \equiv E_0 \hat{n}$ and $m$ is the electron mass \cite{naaman2012chiral}. Therefore, a chiral potential leads to a coupling between the electron's spin and its linear momentum and through this coupling a spin-polarized electron selectively interacts with a chiral molecule. 

To further elucidate this enantioselective interaction, we can consider the energy splitting associated with the motional and spin states of an electron. With respect to a molecular axis an electron can move up ($+$) or down ($-$) and its spin can be +1/2 ($\uparrow$) or -1/2 ($\downarrow$). Therefore we deal with four states, as shown in Figure 2a: $|+,\uparrow\rangle$, $|-,\uparrow\rangle$, $|+,\downarrow\rangle$, $|-,\downarrow\rangle$. By convention we define the direction of the chiral potential such that for an electron moving in the + direction through a right-handed molecule the sign of the magnetic field is positive. Therefore, the interaction of a right-handed molecule with an electron moving in the + direction prefers the -1/2 spin-state of the electron. In other words, -1/2 spin-state is stabilized as it is lower in energy. Similarly, a left-handed molecule preferably interacts with a +1/2 spin-state electron moving in the + direction. Therefore, we can consider doubly degenerate helicity states, $|D\rangle\equiv\left\{|+, \uparrow\rangle, |-, \downarrow\rangle\right\}$ and $|L\rangle\equiv\left\{|+, \downarrow\rangle, |-, \uparrow\rangle\right\}$, for the electron in its interaction with a chiral potential. With this definition, a right-handed molecule energetically prefers $|L\rangle$ electron and vice-versa and the penalty for the helicity flip, $P_{\text{flip}}$, is given by the Boltzmann factor corresponding to the energy gap between $|L\rangle$ and $|D\rangle$ states:

\begin{equation}
    P_{\text{flip}} = \exp\left(-{\frac{2H_{SO}}{k_BT}}\right)
\end{equation}

This factor is responsible for the spin-polarization observed in the CISS experiments as it corresponds to the probability of an electron to be backscattered while retaining its original spin orientation. In order to account for the observed spin-polarizations (at room temperature) one needs to consider a spin-orbit energy, $H_{\text{SO}}$, of around $50$ meV. For larger chiral molecules like double-strand DNA, $H_{\text{SO}}$ can be as large as $500$ meV showing that CISS-like-effects are robust at room temperature \cite{naaman2012chiral}. However, we should emphasize that the $H_{\text{SO}}$ we consider and utilise here is not the actual spin-orbit coupling for the molecule but an effective \textit{ad hoc} value to account for the observations. It is likely that the effective energy, $H_{\text{SO}}$, is due to a combination of spin-orbit and electron spin exchange interactions. The theoretically expected values of spin-orbit coupling for organic molecules are not large enough to explain the observed values of spin polarization. Theory has been catching up recently and it has been proposed that electron-electron interactions can play a significant role in the CISS effect and explain the high degree of spin-polarization for realistic values of spin-orbit coupling \cite{fransson2019chirality}. Using an intrinsic Rashba coupling approach provides some analogous conclusions \cite{varela2019intrinsic}. However, a unifying theoretical description of all CISS-like effects is still an active area of research \cite{evers2021theory}.

\section{Enantioselective Reduction Chemistry}

The CISS effect demonstrates a strong coupling between molecular chirality and the electron spin that is robust at room temperature. This strong coupling suggests that the CISS effect might also be utilized in reverse\textemdash namely, electron spin can bias molecular chirality. In other words, a spin polarized current of electrons can selectively interact with one handedness as the interaction with the opposite handedness is energetically penalized with $P_{\text{flip}}$. However, how exactly is one handedness differentiated from the other in this process?

Relying on the spin-chirality coupling established by the CISS effect, we propose CISS-driven reduction chemistry (CDRC). This CDRC can induce enantiomeric bias, starting from a racemic mixture, due to the selective reaction rates for different chiralities. Consider two reduction reactions, in a solution, that are identical except that one is for an L-molecule (L-A) and the other is for a D-molecule (D-A):

\begin{center}
    \ce{(L/D)-A + e^- -> (L/D)-B-}\\
\end{center}

When the reduction is driven by non-polarized electrons, \ce{e-}, the rates of two reactions are on-average identical and the L-molecule yields an L-product (\ce{L-B-}) and vice versa with the same rate. This is because, on average, half of the electrons are in the right-handed helicity state and they react faster with left-handed molecules and the other half are in the left-handed helicity state and they react faster with the right-handed molecules, therefore the balance is preserved. However, if the helicity of electrons are biased towards one direction, then the chiral selection occurs as one handedness would on-average react faster than the other. The difference in the reaction rates is due to the Arrhenius relation for which the reaction rate, $k$, is proportional to an exponential factor decreasing with higher activation energy: $\exp\left[-\frac{\left(E_a\pm H_{\text{SO}}\right)}{k_BT}\right]$, where $E_a$ is the bare activation energy without the spin-orbit correction. Therefore, the Arrhenius relation predicts a faster reaction rate by $\exp\left(\frac{2H_{\text{SO}}}{k_BT}\right)$ for the `opposite' molecular chirality. For example, for a reduction reaction driven by electrons in the right-handed helicity state, \ce{D-e^-}, we claim that the left-handed reagent, \ce{L-A}, reacts faster:

\begin{center}
    \ce{D-A + D-e^- -> D-B-}: \quad $k_D$\\
    \ce{L-A + D-e^- -> L-B-}: \quad $k_L$\\
\end{center}

where $\frac{k_L}{k_D}=\exp\left(\frac{2H_{\text{SO}}}{k_BT}\right)$ due to the proposed CDRC. This ratio between the reaction rates for different chiralities varies from $5$ to several thousand for effective spin-orbit energies ($20-100$ meV) typically used to account for CISS measurements \cite{naaman2012chiral}. Above, we considered the reduction of a chiral molecule, not the formation of a new chiral center from an achiral precursor (e.g., reduction of \textbf{4} to \textbf{6} in Figure 3.). However, the same effect can still play a role in the latter as soon as there is a chiral reaction intermediate involved in the electron exchange.

For a microscopic explanation of the enantioselectivity in reduction chemistry, we can look into the molecular orbital energies of molecules involved in the electron exchange. Let's first consider a regular redox reaction with non-chiral molecules where the electrons are flowing from a reductant to an oxidant, as in Figure 2b. The driving force for this electron transfer is the energy difference between the HOMO of the reductant and the LUMO of the oxidant and this positive energy difference makes the electron transfer thermodynamically favorable (downhill). However, when the electron acceptor (oxidant) is a chiral molecule and the electron donor (reductant) is ejecting helical electrons, then the LUMO energy of the electron accepting chiral molecule will depend on the molecular handedness. The energy difference between the LUMOs of L and D molecules ($2H_{\text{SO}}$) causes a difference between the reaction rates for two isomers and it is the reason behind the enantioselectivity of the reduction reaction. We can see a manifestation of this in the recent electrochemistry experiments with a magnetized electrode, where the electron spin direction enantioselectively affects the redox behavior of a chiral ferrocene derivative, Ugi's amine, and camphorsulfonic acid \cite{bloom2020asymmetric, metzger2020electron}.

We must note that spin-polarization of electrons \textit{per se} is not enough for enabling CDRC. This is because the CISS effect requires a well defined relationship between the spin and momentum vectors. In fact, chirally induced spin selectivity is the filtering of longitudinally polarized electrons where spin and momentum vectors are co-linear \cite{naaman2012chiral}. Hence, one cannot realize CISS-like effects with spin-polarized electrons with arbitrary momenta. Thermally polarized electrons under very strong magnetic fields or spin-polarized electrons attained by the Triplet Mechanism \cite{atkins1974electron, ces1997spin} are examples where electronic spin polarization is achieved, but helicity is not. This brings up the question of where one could find spin-polarized electrons with well-defined momenta in a prebiotic setting.


\section{Prebiotic Environments, Evaporative Lakes and Magnetite Formation}

The irradiation of a ferromagnetic substrate can provide secondary spin-polarized electrons with an aligned flux of momenta. Due to angular momentum conservation, the spin of electrons ejected from a magnetic surface are aligned. Moreover, the back-scattered photoelectrons move along the surface normal therefore have a well defined momenta. Hence electrons ejected from a magnetized surface have a helical character and their helicity states change depending on the magnetization direction of the substrate (Figure 1).

Such magnetized surfaces\textemdash namely of magnetite\textemdash might have been common on prebiotic early Earth (and early Mars) in the basins of closed evaporative lakes, which also provide environments for surficial origins of life scenarios \cite{patel2015common, sasselov2020origin, damerdeamer2020}. Under anoxic conditions and lake waters rich in dissolved iron, the redox-stratified water column will allow the accumulation of iron oxides deposits\textemdash primarily magnetite (\ce{Fe3O4}) in the deeper layers, below the UV-light photic zone \cite{schrautzerguth1976, klein2005, toscaetal2018}. Evidence for an ancient redox-stratified lake underlied by magnetite-rich sediment was uncovered by the Curiosity rover in Gale crater on Mars \cite{hurowitzetal2017}. Gale crater lake seems to be a good analog for aqueous basins that evolve into environments for origins of life chemistry\textemdash see Fig. 2 in \cite{sasselov2020origin} and their discussion of authigenic magnetite formation. Note that magnetite formation in such lakes would normally precede\textemdash by $10{^3}$-$10{^5}$ years, the prebiotic chemistry pond (see Fig. 2A vs. 2D in \cite{sasselov2020origin}), and hence the flat magnetite surfaces can be exposed to UV light in the shallow photic zone at that later time (Figure 1).

Magnetite has a particularly well-matching work function of around $4$ eV ($310$nm) \cite{huguenin1973photostimulated, eib1975spin} to the anoxic (no ozone) atmosphere of early Earth and the irradiation of a natural magnetite near its photo-threshold has shown to generate spin-polarized electrons with around $-40\%$ polarization \cite{eib1975spin, alvarado1976final}. The low value of the work function of magnetite allows for the generation of photo-electrons for UV wavelengths below $310$nm that were abundant (around two orders of magnitude higher compared to now) on the surface of a prebiotic lake \cite{ranjan2017constraints}. Magnetite is also the most common and naturally magnetic mineral on Earth with a saturation magnetization of $480$ kA/m ($6000$ Gauss) and a Curie temperature of $580^{\circ}$C. Moreover, magnetite maintains magnetization after the external field is removed and the mineral is thermally processed, having a thermoremanent magnetization of $4.8$-$7.2$ kA/m ($60$-$90$ Gauss) \cite{dunlop2005magnetic}. Therefore, within the mesoscopic vicinity of the magnetite deposits there exists a magnetic field stronger than that of the Earth's (around $0.5$ Gauss) field. Individual magnetite sediments in a large area (continent-size to hemisphere) will share the same field orientation. Hence, this homogeneous and relatively strong field can open up the possibility of having magnetic field effects in the aqueous chemistry through spin effects\textemdash especially in regard to photochemical reactions \cite{timmel2004study, hore2020spin}.

\section{Relevance to Prebiotic Chemistry}

\begin{figure}[H]
    \centering
    \includegraphics[scale = 0.8]{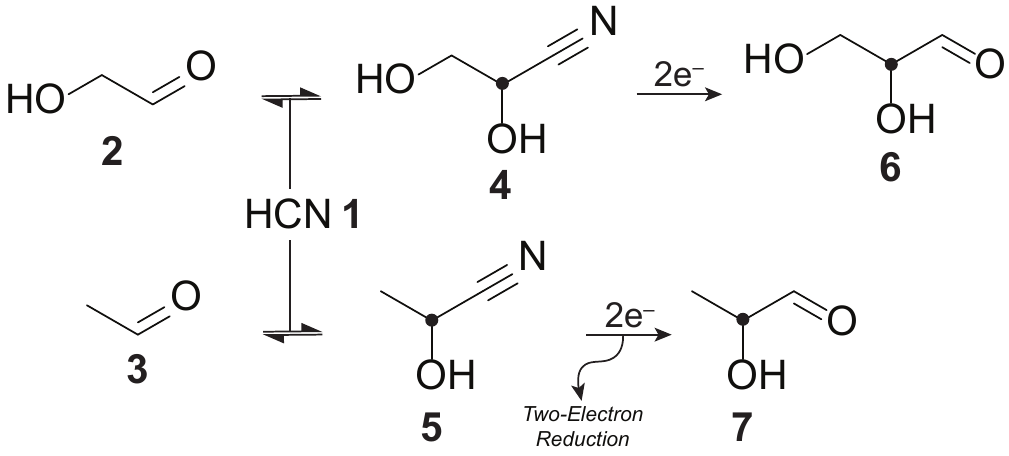}
    \caption{The cyanosulfidic prebiotic chemistry uses solvated electrons as the main reducing agent. The enantioselective reduction scheme we propose can be applied to the cyanosulfidic chemistry when a chiral center is produced and is subjected to a reduction by helical electrons. Black dots show the chiral centers. Adapted from \cite{ritson2013synthesis}.}
    \label{Fig.3}
\end{figure}

We suggest that CDRC with spin-polarized electrons can be realized in prebiotic chemistry networks and an enantiomeric excess can be induced right after the generation of the first chiral molecules. The cyanosulfidic chemistry is an especially attractive candidate as it utilizes solvated electrons for the synthesis of bio-molecules, including the first chiral sugar and chiral precursors of amino acids \cite{xu2018photochemical}. Furthermore, the chemistry is a surface pond chemistry where high-energy UV photons are available for the generation of spin-polarized photoelectrons from a magnetized surface on the shallow lakebed. 

We assume that UV-generated photoelectrons from the magnetite will solvate in water and be utilized effectively in the cyanosulfidic chemistry. Such solvation was demonstrated recently in experiments where photoelectrons ejected from a metallic surface upon UV irradiation become hydrated at the metal/liquid interface \cite{lapointe2020probing}. In the cyanosulfidic chemistry hydrated electrons are generated from the photo-redox cycling of dissolved ferrocyanide and used in the reductive homologation of hydrogen cyanide\textemdash the main feedstock molecule \cite{xu2018photochemical, patel2015common}. The electrons generated in the bulk by photo-redox cycling do not have a helical character, therefore, they do not react enantioselectively. However, we suggest that helical electrons ejected from the magnetite surfaces with a long spin decay time of $8\mu$s \cite{jeevarajan1989esr} can induce enantioselective reduction reactions.

As illustrated in Figure 3, during the first reduction stage, glycolaldehyde \textbf{2} and acetaldehyde \textbf{3} are synthesized from glycolonitrile. Following this step, glycolaldehyde cyanohydrin \textbf{4} and acetaldehyde cyanohydrin \textbf{5} are yielded by a Killiani-Fischer type growth of glycolaldehyde and acetaldehyde with hydrogen cyanide \textbf{1}, respectively \cite{ritson2013synthesis}. These cyanohydrins are the first chiral molecules in the cyanosulfidic chemistry and they undergo a prompt reduction with solvated electrons to produce glyceraldehyde \textbf{6} (first chiral sugar) and lactaldehyde \textbf{7} (threonine precursor). This is the step where CDRC can play a role as we consider the reaction of a chiral molecule with an electron. \textit{If electrons with defined helicity are used to induce this reduction of chiral cyanohydrins we suggest that an ee can be induced in the synthesis of glyceraldehyde and aminoacid precursors.} 

Given that glyceraldehyde \textbf{6} is also shown to play a vital role in the Powner-Sutherland-type synthesis of nucleotides \cite{powner2009synthesis}, inducing an ee at the early stages of the prebiotic chemistry network is significant, as it carries on at the later stages. Having a persistent chiral bias increases the yield of each step and this becomes especially important in such multi-step reaction networks where the overall yield can be quickly converged to zero by the so-called \textit{arithmetic demon} \cite{crispino1993selective, rimmer2018origin}. 

In addition, the effects of photo-oxidation on the magnetite surface can be reversed and the surface can be refreshed by reductants such as bisulfite (\ce{HSO3-}) which has also been shown to concentrate and stabilize aldehydes, and prevent the racemization of glyceraldehyde into dihydroxyacetone \cite{ritson2018mimicking}.

Note that the enantioselective effect we propose can be considered in combination with a dynamical kinetic resolution scheme that can amplify induced ee. If helical electrons can indeed lead to a selectivity in the reaction rates for different enantiomers at the reduction stage (e.g. \textbf{4} to \textbf{6}), dynamic resolution can enhance the ee at the product stage (\textbf{6}) to more than $50\%$ due to racemization ($\textbf{2} \rightleftharpoons \textbf{4}$) of the reagent (\textbf{4}).


\section{Discussion}

Let us summarize the features of CDRC as a potential mechanism for inducing ee in prebiotic chemistry and therefore being the trigger of biological homochirality.

The basic mechanism behind CDRC is the robust coupling between molecular chirality and the electron spin established by the CISS effect. As such, CDRC is a robust and deterministic way of inducing ee at room temperature and in solution.

Moreover, the mechanism we suggest is prebiotically plausible. Helical photo-electrons can be generated by irradiating magnetized magnetite deposits with UV light. Therefore, the chiral symmetry breaking agents\textemdash the helical electrons\textemdash are generated \textit{in situ} with the bio-molecules and their supply is robust and long-lasting. Magnetite minerals and high-energy UV radiation capable of ejecting photo-electrons are expected to be available in prebiotic lake environments. 

CDRC takes advantage of the exponential decrease in the reaction rates with the activation energy. Therefore, even for small chiral molecules with effective spin-orbit couplings of around $50$ meV it is possible to achieve a $50$-fold difference in the reaction rates for the two enantiomers.
Furthermore, our mechanism induces ee by means of synthesis, not by enantioselective destruction. Therefore, it does not deteriorate the reaction yields in a multi-step reaction network. This adds to the increasing reaction yields in the approach to a chirally purer mixture and boosts the prebiotic plausibility of the entire network. 

The fundamental coupling at the core of the CDRC is compatible with any chiral molecule as long as it reacts with low-energy, spin-polarized electrons. Hence, it would apply to various reaction steps where a chiral molecule is reduced, in any prebiotic chemistry scenario. For a network, of course, the earlier the better, as it will contribute to higher yields. 

As prebiotic chemistry scenarios go, CDRC seems best applicable to the cyanosulfidic chemistry network \cite{xu2018photochemical, patel2015common}, because UV-generated solvated electrons are essential in the initial synthesis. Starting with the second-stage reduction of nitriles in the cyanosulfidic network, one can utilize helical photoelectrons to induce ee on (or near) the surface while non-helical solvated electrons in the bulk still react in a chirally non-selective way. Therefore, in the prebiotic scenario we conceive, surface electrons induce ee via CDRC and the bulk electrons, generated by photo-redox cycling, keep acting as the main reducing agent. Although the non-helical bulk electrons are likely to outnumber the surface electrons due to the low efficiency ($10^{-7}$-$10^{-2}$ depending on the photon energy \cite{huguenin1973photostimulated}) of backscattering in the photoelectric effect, ee can still be induced near the surface. However, because we do not yet know the yield-limiting step of the whole process, it is hard to accurately estimate the helical solvated electron yield and the subsequent efficiency of inducing ee. We, therefore, intend to address the yields question with future experimental work.

Another remaining question is if the helical electrons will remain helical when they are solvated. This is an important step, because when photoelectrons are ejected from the magnetized surface they have a finite length scale before their helicity is lost. Although it is hard to predict the hydration dynamics of a spin-polarized electron and estimate the lifetime of the helical electrons, the long spin-relaxation time of solvated electrons (T\textsubscript{1}$\approx8\mu s$ \cite{jeevarajan1989esr}) and the observation of the spin-dependent electrochemical behavior with magnetized electrodes are encouraging \cite{bloom2020asymmetric, metzger2020electron}. We hope that future experiments under study are going to elucidate the validity of the aforementioned claims.

A possible experiment we conceive of is irradiating a magnetized magnetite surface with UV light in the presence of a racemic mixture of a chiral electron acceptor. Measuring the magnitude of the induced ee, if any, of the chiral reduction product over time can reveal the occurrence of the chiral symmetry breaking and subsequent amplification of the ee.


\section{Conclusion}

This work offers a prebiotically plausible mechanism for elucidating the origin of life's homochirality. We propose spin-polarized electrons ejected from magnetized magnetite deposits by UV irradiation as likely candidates for breaking the chiral symmetry in prebiotic chemistry. The suggested mechanism relies on the interaction between electron spin and molecular chirality\textemdash whose magnitude is empirically derived from the CISS effect\textemdash and it anticipates different synthetic reaction rates for the enantiomers. Thereby, the reduction with helical electrons can break the chiral symmetry and pave the way for enriching the chiral excess over time where the enhancement rate is determined by the difference in the reaction rates for specific enantiomers. The mechanism is robust at room temperature, in solution, and is applicable to any chiral structure. It offers an \textit{in situ} and continuous generation of enantiomeric excess in a prebiotic lake environment. Moreover, the generation of chiral excess with spin-polarized electrons seems well-suited to surficial prebiotic chemistry scenarios, like the cyanosulfidic chemistry, which relies on solvated secondary electrons as its main reducing agent. We expect that future experiments can test the mechanism's viability.

\section*{Acknowledgments}

The authors thank Oren Ben-Dor, Donna Blackmond, Corinna Kufner, Vladimiro Mujica, Ron Naaman, John Sutherland, and Jack Szostak for helpful discussions, suggestions, and feedback. The authors further thank Oren Ben-Dor, Donna Blackmond, and Jack Szostak for their review of the draft of this work. We also acknowledge other members of the Simons Collaboration on the Origins of Life and the Harvard Origins of Life Initiative for fruitful discussions that shaped the ideas behind this work. This work was supported by a grant from the Simons Foundation 290360 to D.D.S.

\printbibliography

@article{joyce1984chiral,
  title={Chiral selection in poly (C)-directed synthesis of oligo (G)},
  author={Joyce, GF and Visser, GM and Van Boeckel, CAA and Van Boom, JH and Orgel, LE and Van Westrenen, J},
  journal={Nature},
  volume={310},
  number={5978},
  pages={602--604},
  year={1984},
  publisher={Nature Publishing Group}
}

@article{blackmond2010origin,
  title={The origin of biological homochirality},
  author={Blackmond, Donna G},
  journal={Cold Spring Harbor perspectives in biology},
  volume={2},
  number={5},
  pages={a002147},
  year={2010},
  publisher={Cold Spring Harbor Lab}
}

@article{globus2020chiral,
  title={The chiral puzzle of life},
  author={Globus, Noemie and Blandford, Roger D},
  journal={The Astrophysical Journal Letters},
  volume={895},
  number={1},
  pages={L11},
  year={2020},
  publisher={IOP Publishing}
}

@article{rikken2000enantioselective,
  title={Enantioselective magnetochiral photochemistry},
  author={Rikken, GLJA and Raupach, E},
  journal={Nature},
  volume={405},
  number={6789},
  pages={932--935},
  year={2000},
  publisher={Nature Publishing Group}
}

@book{kelvin1904baltimore,
  title={Baltimore lectures on molecular dynamics and the wave theory of light},
  author={Kelvin, Lord William Thomson},
  year={1904},
  publisher={CUP Archive}
}

@article{ranjan2017constraints,
  title={Constraints on the early terrestrial surface UV environment relevant to prebiotic chemistry},
  author={Ranjan, Sukrit and Sasselov, Dimitar D},
  journal={Astrobiology},
  volume={17},
  number={3},
  pages={169--204},
  year={2017},
  publisher={Mary Ann Liebert, Inc. 140 Huguenot Street, 3rd Floor New Rochelle, NY 10801 USA}
}

@article{sasselov2020origin,
  title={The origin of life as a planetary phenomenon},
  author={Sasselov, Dimitar D and Grotzinger, John P and Sutherland, John D},
  journal={Science Advances},
  volume={6},
  number={6},
  pages={eaax3419},
  year={2020},
  publisher={American Association for the Advancement of Science}
}

@article{xu2018photochemical,
  title={Photochemical reductive homologation of hydrogen cyanide using sulfite and ferrocyanide},
  author={Xu, Jianfeng and Ritson, Dougal J and Ranjan, Sukrit and Todd, Zoe R and Sasselov, Dimitar D and Sutherland, John D},
  journal={Chemical Communications},
  volume={54},
  number={44},
  pages={5566--5569},
  year={2018},
  publisher={Royal Society of Chemistry}
}

@article{powner2009synthesis,
  title={Synthesis of activated pyrimidine ribonucleotides in prebiotically plausible conditions},
  author={Powner, Matthew W and Gerland, B{\'e}atrice and Sutherland, John D},
  journal={Nature},
  volume={459},
  number={7244},
  pages={239--242},
  year={2009},
  publisher={Nature Publishing Group}
}

@article{ritson2013synthesis,
  title={Synthesis of aldehydic ribonucleotide and amino acid precursors by photoredox chemistry},
  author={Ritson, Dougal J and Sutherland, John D},
  journal={Angewandte Chemie International Edition},
  volume={52},
  number={22},
  pages={5845--5847},
  year={2013},
  publisher={Wiley Online Library}
}

@article{rimmer2018origin,
  title={The origin of RNA precursors on exoplanets},
  author={Rimmer, Paul B and Xu, Jianfeng and Thompson, Samantha J and Gillen, Ed and Sutherland, John D and Queloz, Didier},
  journal={Science advances},
  volume={4},
  number={8},
  pages={eaar3302},
  year={2018},
  publisher={American Association for the Advancement of Science}
}

@article{crispino1993selective,
  title={Selective perhydroxylation of squalene: taming the arithmetic demon},
  author={Crispino, Gerard A and Ho, Pui Tong and Sharpless, K Barry},
  journal={Science},
  volume={259},
  number={5091},
  pages={64--66},
  year={1993},
  publisher={American Association for the Advancement of Science}
}

@article{patel2015common,
  title={Common origins of RNA, protein and lipid precursors in a cyanosulfidic protometabolism},
  author={Patel, Bhavesh H and Percivalle, Claudia and Ritson, Dougal J and Duffy, Colm D and Sutherland, John D},
  journal={Nature chemistry},
  volume={7},
  number={4},
  pages={301--307},
  year={2015},
  publisher={Nature Publishing Group}
}

@article{atkins1974electron,
  title={Electron spin polarization in a rotating triplet},
  author={Atkins, P.W. and Evans, G.T.},
  journal={Molecular Physics},
  volume={27},
  number={6},
  pages={1633--1644},
  year={1974},
  publisher={Taylor \& Francis}
}

@article{ces1997spin,
  title={Spin polarization in triplet states in solution},
  author={Ces, O and McLauchlan, KA and Qureshi, TJJ},
  journal={Applied Magnetic Resonance},
  volume={13},
  number={3},
  pages={297--315},
  year={1997},
  publisher={Springer}
}

@article{jeevarajan1989esr,
  title={ESR studies of solvated electron in liquid solution using photolytic production},
  author={Jeevarajan, AS and Fessenden, RW},
  journal={The Journal of Physical Chemistry},
  volume={93},
  number={9},
  pages={3511--3514},
  year={1989},
  publisher={ACS Publications}
}

@article{shibata1998amplification,
  title={Amplification of a slight enantiomeric imbalance in molecules based on asymmetric autocatalysis: the first correlation between high enantiomeric enrichment in a chiral molecule and circularly polarized light},
  author={Shibata, Takanori and Yamamoto, Jun and Matsumoto, Naoko and Yonekubo, Shigeru and Osanai, Shunji and Soai, Kenso},
  journal={Journal of the American Chemical Society},
  volume={120},
  number={46},
  pages={12157--12158},
  year={1998},
  publisher={ACS Publications}
}

@article{blackmond2019autocatalytic,
  title={Autocatalytic models for the origin of biological homochirality},
  author={Blackmond, Donna G},
  journal={Chemical reviews},
  volume={120},
  number={11},
  pages={4831--4847},
  year={2019},
  publisher={ACS Publications}
}

@article{noorduin2009complete,
  title={Complete chiral symmetry breaking of an amino acid derivative directed by circularly polarized light},
  author={Noorduin, Wim L and Bode, Arno AC and Van Der Meijden, Maarten and Meekes, Hugo and Van Etteger, Albert F and Van Enckevort, Willem JP and Christianen, Peter CM and Kaptein, Bernard and Kellogg, Richard M and Rasing, Theo and others},
  journal={Nature Chemistry},
  volume={1},
  number={9},
  pages={729--732},
  year={2009},
  publisher={Nature Publishing Group}
}

@article{szabo1999demonstration,
  title={Demonstration of the parity-violating energy difference between enantiomers},
  author={Szab{\'o}-Nagy, Andrea and Keszthelyi, Lajos},
  journal={Proceedings of the National Academy of Sciences},
  volume={96},
  number={8},
  pages={4252--4255},
  year={1999},
  publisher={National Acad Sciences}
}

@article{mason1987universal,
  title={Universal dissymmetry and the origin of biomolecular chirality},
  author={Mason, Stephen F},
  journal={Biosystems},
  volume={20},
  number={1},
  pages={27--35},
  year={1987},
  publisher={Elsevier}
}

@article{gohler2011spin,
  title={Spin selectivity in electron transmission through self-assembled monolayers of double-stranded DNA},
  author={G{\"o}hler, B and Hamelbeck, V and Markus, TZ and Kettner, M and Hanne, GF and Vager, Zeev and Naaman, Ron and Zacharias, H},
  journal={Science},
  volume={331},
  number={6019},
  pages={894--897},
  year={2011},
  publisher={American Association for the Advancement of Science}
}

@article{naaman2012chiral,
  title={Chiral-induced spin selectivity effect},
  author={Naaman, Ron and Waldeck, David H},
  journal={The journal of physical chemistry letters},
  volume={3},
  number={16},
  pages={2178--2187},
  year={2012},
  publisher={ACS Publications}
}

@article{rosenberg2008chiral,
  title={Chiral-selective chemistry induced by spin-polarized secondary electrons from a magnetic substrate},
  author={Rosenberg, RA and Haija, M Abu and Ryan, PJ},
  journal={Physical review letters},
  volume={101},
  number={17},
  pages={178301},
  year={2008},
  publisher={APS}
}

@article{rosenberg2015chiral,
  title={Chiral Selective Chemistry Induced by Natural Selection of Spin-Polarized Electrons},
  author={Rosenberg, Richard A and Mishra, Debabrata and Naaman, Ron},
  journal={Angewandte Chemie},
  volume={127},
  number={25},
  pages={7403--7406},
  year={2015},
  publisher={Wiley Online Library}
}

@article{metzger2020electron,
  title={The electron spin as a chiral reagent},
  author={Metzger, Tzuriel S and Mishra, Suryakant and Bloom, Brian P and Goren, Naama and Neubauer, Avner and Shmul, Guy and Wei, Jimeng and Yochelis, Shira and Tassinari, Francesco and Fontanesi, Claudio and others},
  journal={Angewandte Chemie},
  volume={132},
  number={4},
  pages={1670--1675},
  year={2020},
  publisher={Wiley Online Library}
}

@article{huguenin1973photostimulated,
  title={Photostimulated oxidation of magnetite: 2. Mechanism},
  author={Huguenin, Robert L},
  journal={Journal of Geophysical Research},
  volume={78},
  number={35},
  pages={8495--8506},
  year={1973},
  publisher={Wiley Online Library}
}

@article{dunlop2005magnetic,
  title={Magnetic minerals in the Martian crust},
  author={Dunlop, David J and Arkani-Hamed, Jafar},
  journal={Journal of Geophysical Research: Planets},
  volume={110},
  number={E12},
  year={2005},
  publisher={Wiley Online Library}
}

@article{hein2011route,
  title={A route to enantiopure RNA precursors from nearly racemic starting materials},
  author={Hein, Jason E and Tse, Eric and Blackmond, Donna G},
  journal={Nature chemistry},
  volume={3},
  number={9},
  pages={704--706},
  year={2011},
  publisher={Nature Publishing Group}
}

@article{fransson2019chirality,
  title={Chirality-induced spin selectivity: The role of electron correlations},
  author={Fransson, Jonas},
  journal={The journal of physical chemistry letters},
  volume={10},
  number={22},
  pages={7126--7132},
  year={2019},
  publisher={ACS Publications}
}

@article{klein2005,
  title={Some Precambrian banded iron formations (BIFs) from around the world: Their age, geologic setting, minerology, metamorphism, geochemistry and origin},
  author={Klein, Cornelis},
  journal={American Mineralogist},
  volume={90},
  number={},
  pages={1473--1499},
  year={2005},
  publisher={}
}

@article{schrautzerguth1976,
  title={Hydrogen Evolving Systems.I. The Formation of H2 from Aqueous Suspensions of Fe(OH)2 and Reactions with Reducible Substrates, Including Molecular Nitrogen},
  author={Schrautzer, G. and Guth, T.},
  journal={Journal of the American Chemical Society},
  volume={98},
  number={12},
  pages={3508--3513},
  year={1976},
  publisher={ACS Publications}
}

@article{toscaetal2018,
  title={Magnetite Authigenesis and the Warming of Early Mars},
  author={Tosca, Nicolas and Ahmed, Imad and Tutolo, Benjamin and Ashpitel, Alice and Hurowitz, Joel},
  journal={Nature Geoscience},
  volume={11},
  number={9},
  pages={635--639},
  year={2018},
  publisher={}
}

@article{hurowitzetal2017,
  title={Redox stratification of an ancient lake in Gale crater, Mars},
  author={Hurowitz, Joel and Grotzinger, John and Fischer, Woodward and McLennan, S. and et al.},
  journal={Science},
  volume={356},
  number={},
  pages={eaah6849},
  year={2017},
  publisher={American Association for the Advancement of Science}
}

@article{damerdeamer2020,
  title={The Hot Spring Hypothesis for an Origin of Life},
  author={Damer, Bruce and Deamer, David},
  journal={Astrobiology},
  volume={20},
  number={3},
  pages={},
  year={2020},
  publisher={American Association for the Advancement of Science}
}

@misc{hore2020spin,
  title={Spin chemistry},
  author={Hore, PJ and Ivanov, Konstantin L and Wasielewski, Michael R},
  journal={The Journal of chemical physics},
  volume={152},
  number={12},
  pages={120401},
  year={2020},
  publisher={AIP Publishing LLC}
}

@article{timmel2004study,
  title={A study of spin chemistry in weak magnetic fields},
  author={Timmel, Christiane R and Henbest, Kevin B},
  journal={Philosophical Transactions of the Royal Society of London. Series A: Mathematical, Physical and Engineering Sciences},
  volume={362},
  number={1825},
  pages={2573--2589},
  year={2004},
  publisher={The Royal Society}
}

@article{evers2021theory,
  title={Theory of Chirality Induced Spin Selectivity: Progress and Challenges},
  author={Evers, Ferdinand and Aharony, Amnon and Bar-Gill, Nir and Entin-Wohlman, Ora and Hedeg{\aa}rd, Per and Hod, Oded and Jelinek, Pavel and Kamieniarz, Grzegorz and Lemeshko, Mikhail and Michaeli, Karen and others},
  journal={arXiv preprint arXiv:2108.09998},
  year={2021}
}

@inproceedings{eib1975spin,
  title={Spin polarized photoemission from Fe3O4},
  author={Eib, W and Meier, F and Pierce, DT and Sattler, K and Siegmann, HC},
  booktitle={AIP Conference Proceedings},
  volume={24},
  number={1},
  pages={88--89},
  year={1975},
  organization={American Institute of Physics}
}

@article{alvarado1976final,
  title={Final-state effects in the 3 d photoelectron spectrum of Fe 3 O 4 and comparison with Fe x O},
  author={Alvarado, SF and Erbudak, M and Munz, P},
  journal={Physical Review B},
  volume={14},
  number={7},
  pages={2740},
  year={1976},
  publisher={APS}
}

@article{dreiling2014chirally,
  title={Chirally sensitive electron-induced molecular breakup and the Vester-Ulbricht hypothesis},
  author={Dreiling, JM and Gay, Timothy J},
  journal={Physical review letters},
  volume={113},
  number={11},
  pages={118103},
  year={2014},
  publisher={APS}
}

@article{ray1999asymmetric,
  title={Asymmetric scattering of polarized electrons by organized organic films of chiral molecules},
  author={Ray, K and Ananthavel, SP and Waldeck, DH and Naaman, Ron},
  journal={Science},
  volume={283},
  number={5403},
  pages={814--816},
  year={1999},
  publisher={American Association for the Advancement of Science}
}

@article{naaman2019chiral,
  title={Chiral molecules and the electron spin},
  author={Naaman, Ron and Paltiel, Yossi and Waldeck, David H},
  journal={Nature Reviews Chemistry},
  volume={3},
  number={4},
  pages={250--260},
  year={2019},
  publisher={Nature Publishing Group}
}

@article{lapointe2020probing,
  title={Probing the birth and ultrafast dynamics of hydrated electrons at the gold/liquid water interface via an optoelectronic approach},
  author={Lapointe, François and Wolf, Martin and Campen, R Kramer and Tong, Yujin},
  journal={Journal of the American Chemical Society},
  volume={142},
  number={43},
  pages={18619--18627},
  year={2020},
  publisher={ACS Publications}
}

@article{ritson2018mimicking,
  title={Mimicking the surface and prebiotic chemistry of early Earth using flow chemistry},
  author={Ritson, Dougal J and Battilocchio, Claudio and Ley, Steven V and Sutherland, John D},
  journal={Nature communications},
  volume={9},
  number={1},
  pages={1--10},
  year={2018},
  publisher={Nature Publishing Group}
}

@article{varela2019intrinsic,
  title={Intrinsic Rashba coupling due to hydrogen bonding in DNA},
  author={Varela, S and Monta{\~n}es, B and L{\'o}pez, F and Berche, B and Guillot, B and Mujica, V and Medina, E},
  journal={The Journal of Chemical Physics},
  volume={151},
  number={12},
  pages={125102},
  year={2019},
  publisher={AIP Publishing LLC}
}

@article{shibata1999practically,
  title={Practically perfect asymmetric autocatalysis with (2-Alkynyl-5-pyrimidyl) alkanols},
  author={Shibata, Takanori and Yonekubo, Shigeru and Soai, Kenso},
  journal={Angewandte Chemie International Edition},
  volume={38},
  number={5},
  pages={659--661},
  year={1999},
  publisher={Wiley Online Library}
}

@article{legnani2021mechanistic,
  title={Mechanistic Insight into the Origin of Stereoselectivity in the Ribose-Mediated Strecker Synthesis of Alanine},
  author={Legnani, Luca and Daru, Andrea and Jones, Alexander X and Blackmond, Donna G},
  journal={Journal of the American Chemical Society},
  volume={143},
  number={20},
  pages={7852--7858},
  year={2021},
  publisher={ACS Publications}
}

@article{balavoine1974preparation,
  title={Preparation of chiral compounds with high optical purity by irradiation with circularly polarized light, a model reaction for the prebiotic generation of optical activity},
  author={Balavoine, G and Moradpour, A and Kagan, HB},
  journal={Journal of the American Chemical Society},
  volume={96},
  number={16},
  pages={5152--5158},
  year={1974},
  publisher={ACS Publications}
}

@article{bloom2020asymmetric,
  title={Asymmetric reactions induced by electron spin polarization},
  author={Bloom, BP and Lu, Y and Metzger, Tzuriel and Yochelis, Shira and Paltiel, Yossi and Fontanesi, Claudio and Mishra, Suryakant and Tassinari, Francesco and Naaman, Ron and Waldeck, DH},
  journal={Physical Chemistry Chemical Physics},
  volume={22},
  number={38},
  pages={21570--21582},
  year={2020},
  publisher={Royal Society of Chemistry}
}
\end{document}